\documentclass[12pt]{article}
\usepackage{graphics,epsfig}
 
\begin{document}
\title{From 2000 Bush-Gore to 2006 Italian elections: Voting at fifty-fifty and the Contrarian Effect}
\author{Serge Galam\\
 Centre de Recherche en \'Epist\'emologie Appliqu\'ee (CREA),\\
\'Ecole Polytechnique et CNRS (UMR 7656),\\
1,  rue Descartes, 75005 Paris, France}
\date{serge.galam@polytechnique.edu}
\maketitle

\begin{abstract}

A sociophysical model for opinion dynamics is shown to embody a series of recent western hung national votes all set at the unexpected and very improbable edge of a fifty-fifty score. It started with the Bush-Gore 2000 American presidential election, followed by the 2002 Stoiber-Schr\H{o}der, then the 2005 Schr\H{o}der-Merkel German elections, and finally the 2006 Prodi-Berlusconi Italian elections. In each case, the country was facing drastic choices, the running competing parties were advocating very different programs and millions of voters were involved. Moreover, polls were given a substantial margin for the predicted winner. While all these events were  perceived as accidental and isolated, our model suggests that indeed they are deterministic and obey to one single universal phenomena associated to the effect of contrarian behavior on the dynamics of opinion forming. The not hung Bush-Kerry 2005 presidential election is shown to belong to the same universal frame. To conclude, the existence of contrarians hints at the repetition of hung elections in the near future.

\end{abstract}


\section{Voting at fifty-fifty: a very rare accidental event ?}

The mere principle underlying a democratic voting is the  assumption that there always exists a majority, even down to a small margin of a few percents. Voting at fifty-fifty sounds an impossible event both statistically and socially, in particular when millions of voters are involved. The impossibility looks total when simultaneously the country is facing a difficult situation with drastic political decisions at stake, and both competing parties are advocating radical differences in their respective programs. 

One such typical case was the recent Italian general elections of April 2006, which were predicted to mark a solid victory of Prodi coalition at the expense of the  sitting coalition of Berlusconi. Exit polls did confirmed this expectation before  the actual results dismissed it by establishing a fifty-fifty score in terms of votes. Indeed Prodi coalition won at a margin of 24 755 votes, i.e., a difference of 0.07$\%$, out of  more than 38 million votes cast \cite{wiki}.

Such a hung election scenario was a surprising and unexpected rare event for a national election where the two competing coalitions had so different political program to address  a controversial assessment of the Berlusconi ending term policy. But indeed, it happened already several times in few western democracies starting from the year 2000 in the US with the famous Bush-Gore stand up for the swing state of Florida. it eventually ends with the victory in favor of Bush by an excess of 537 votes. In terms of national votes Bush got 47.9$\%$ against 48.4$\%$ for Gore yielding a margin of 0.5$\%$ of the total 105 417 258 votes \cite{wiki}.

A similar scenario occurred again and twice in Germany for the federal elections, first in  2002 with the Stoiber-Schr\H{o}der race, and then in 2005 with the Schr\H{o}der-Merkel race, In the last case Merkel CDU/CSU alliance won  35.2 $\%$ of the votes against 34.2 $\%$ for Schr\H{o}der SPD scoring a difference of 0.92 $\%$. The novelty from this vote with respect to the previous September 2002 one,  was the fact that the SPD/Green coalition fell to reach a majority at the parliament in term of seats. In 2002, although the CDU/CSU and SPD both got 38.5$\%$ of the votes cast, the SPD and Green reached 50.7$\%$  in seat percentage \cite{wiki}. 

The 2005 Bush-Kerry American presidential elections is at odd with above cases since Bush received a majority of 50.73 $\%$ to 48.27$\%$ for Kerry making a 2.5$\%$ margin. Nevertheless, it subscribes to the same scenario as will become clear latter. We claim that it was the Bin Laden videotape released the Friday preceding the election which shifted the election vote from a hung result. Accordingly, would the elections have taken place two weeks latter, the Bush-Kerry race would have yield a hung election as in 2000 as seen from the polls evolution during the campaign.

These very rare events all came as a tremendous surprise  and were not given much explanation beside being accidental. They were neither associated to a single social phenomena. At contrast, here we want to address the question of the possible origin and connexion of above series of recent hung elections. In particular to find out if they could be reproduced using one single new universal feature to be included in the dynamics of opinion forming. And subsequently to determine if indeed  the hung scenario can be the result of a deterministic mechanism.

\section{Sociophysics and opinion dynamics}

To address above basic questions a general frame for opinion dynamics is given in terms of a simple sociophysical model which operates using local interactions and iterative updates of individual opinions \cite{chopard1}. The model combines rational individual choices \cite{mosco1} and some contrarian behavior \cite{contra}. A contrarian is someone who deliberately decides to oppose the prevailing choice of others whatever this choice is \cite{contra,contra-eco}. 

It shows that a public debate leads to the emergence of a stable majority. The associated dynamics is monitored by one separator located at fifty-fifty percent and two symmetric attractors featuring a stable coexistence of a clear cut majority-minority. It is the initial majority of individual choices prior to the campaign, which eventually wins the vote. The level of the minority is function of the density of contrarians and the time duration of the public debate.

However, it is found that the existence of contrarians, above some density, reverses  the whole dynamics with the sudden merging of above two attractors with the  separator, which in turn results in one single attractor located exactly at a fifty-fifty support.  

Within our model of opinion dynamics a coherent and single light  is thus shed to these very rare and disconected hung election events which occurred recently in several western democracies. They appear to be all embodied in a single frame and produced by the unique universal feature of contrarian behavior. Such an explanation hints at the repetition of hung elections in the near future.

This work contributes to the now growing field of applications of 
statistical physics to opinion dynamics  
\cite{sorin,mino,weron,espagnol,caruso} as part
of the wider field of  ``sociophysics"  \cite{strike}, which studies collective social and political behaviors. We stress that we are not aiming at an exact description of the real social
and political life, but rather, doing some crude approximations, to enlighten
essential features of an otherwise very complex and multiple phenomena. 

It is worth to emphasize that in 2005, for the first time, a highly improbable political vote outcome was predicted  using our cultural bias model of sociophysics \cite{lehir}. The prediction was made several months ahead of the actual vote against all polls and analyses predictions. The model deals with the dynamics of spreading of a minority opinion in public debates using a two sate variable system  \cite{mino,hetero}. It applies to a large spectrum of issues including national votes like the recent French vote, behavior changes like smoking versus non-smoking, support or opposition to a military action like the war in Iraq, rumors like the French hoax about September eleven \cite{rumor}, and reform proposals \cite{lemonde}. 

The emergence of a stable collective opinion is found to obey a threshold dynamic.  For  general issues it is the degree of heterogeneity in the distribution of common beliefs which determines the current value of the threshold which may well vary from ten percents to ninety percents. Accordingly the expected democratic character of a free public debate may turn  onto a dictatorial machine to propagate the opinion of a tiny minority against the initial opinion of the overwhelming majority  \cite{hetero}.

\section{Our model of opinion dynamics}

To study the opinion dynamics of two competing choices we discriminate between two levels in the process of formation of an individual choice. The first one is external and sums up the net result from respectively the global information available to every one, the private information some persons may have, the individual beliefs, and the influence of mass media. The second level is internal and concerns the dynamics of interaction driven by people discussions. In real life although both levels are independent, active and interpenetrated simultaneously, here to study specifically the laws governing the internal dynamics resulting from people discussions, we arbitrarily decoupled them without loss of generality \cite{mino,contra}.

Along this scheme we consider a population of N agents prior to the starting of the public debate. At a time $t$ we measure the opinion distribution which results from the external level by polling the respective support for each of the two competing coalitions denoted A and B.  From the numbers $N_A(t)$ and $N_B(t)$ of agents having the intention to vote for each coalition we obtain the associated individual  probabilities at time $t$
\begin{equation}
p_{A,B}(t)\equiv \frac{N_{A,B}(t)}{N}  ,
\end{equation} 
with,
\begin{equation}
p_A(t)+p_B(t)=1 ,
\end{equation}
where we assumed every agent do have an opinion with $N_A(t)+N_B(t)=N$. From an initial configuration, we consider artificially that all external effects are cut off at time $t$ and then the public debate is activated with people starting to meet and argue among themselves. 

Of course,
they don't meet all the time and all together at once. These sociopolitical gatherings are shaped by the geometry of social life within physical spaces like offices, houses, 
bars, restaurants and others. This geometry determines the times and the numbers of people, which meet at a given place. Usually it is of the order of just a few. Groups may be larger but in these cases spontaneous splitting always occurs with people discussing in smaller subgroups.  The social meeting times are distributed around daily highlights like lunches, dinners and drinks.

To emphasize the  cognitive mechanism at work in the  opinion dynamics which arises 
from local interactions within each social group meeting, we use a simple local majority rule. The complicated, and yet mainly unknown psychological process of an individual mind shift, driven by a local group discussions, is thus implemented by a simple  ``one person - one argument'' principle. 

No advantage is given to a coalition in terms of 
lobbying or organized strategy. People only discuss in small groups with an identical individual persuasive power.  At the end of a meeting all members of a group adopt the opinion which had the initial majority of given arguments. In case there exists no majority, i.e. at a tie in a group of even size, each members preserves its initial opinion.

\begin{figure}[t]
\includegraphics[width=.75\textwidth]{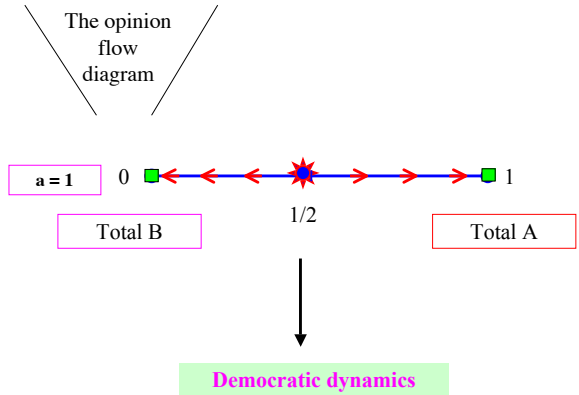}
\caption{
Variation of the flow diagram of opinion dynamics when there exists no contrarians $(a=1)$ for update groups of size $3$. Arrows show the direction of the flow driven by the public debate for any initial respective supports.}
\label{flow1}
\end{figure}

At the first step of our modeling we restrict the social discussion meeting of agents to groups of size three. This constraint will be released in a second step. Accordingly, in the first cycle of local discussions, groups of size 3 are formed randomly. There, all participants adopt 
the local majority state. An initial 2 A (B) with one B (A) ends up with 3 A (B). The resulting voting probability  intention is
\begin{equation}
p_A(t+1)=p_A(t)^3+3p_A(t)^2p_B(t) ,
\label{p3}
\end{equation}
where $p_A(t+1)>p_A(t)$  if $p_A(t)>\frac{1}{2}$ and $p_A(t+1)<p_A(t)$ 
if $p_A(t)<\frac{1}{2}$ since Eq. (\ref{p3}) makes the probability 
vote intention $p_A(t)$  to flow 
monotonically 
toward either one of two stable point attractors located at respectively $P_{A}=1$ and $P_{B}=0$. 
An unstable point 
separator at $p_c=\frac{1}{2}$ monitors the direction of the opinion flow. It separates the two basins 
of attraction 
associated with the two point attractors as seen in Figure (\ref{flow1})

\begin{figure}[t]
\includegraphics[width=.75\textwidth]{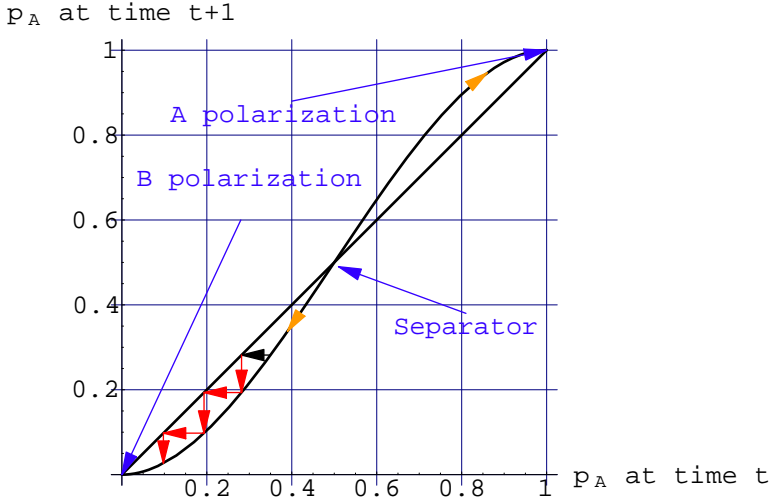}
\caption{
Variation of the proportion $p_A(t+1)$ of supporters as function of $p_A(t)$ for groups of size $3$. Arrows show the direction of the flow for an initial support $p_A(t)<p_{c,3}=\frac{1}{2}$. The flow is democratic.}
\label{p3t}
\end{figure}

During an election campaign people go trough several successive different local 
discussions. To follow the associated vote intention evolution we iterate Eq. (\ref{p3}). A number 
of m discussion cycles gives the series $p_A(t+1), p_A(t+2)... p_A(t+m)$. For instance 
starting at $p_A(t)=0.45$  leads successively after 5 vote intention updates to the series 
$p_A(t+1)=0.43, p_A(t+2)=0.39, p_A(t+3)=0.34, p_A(t+4)=0.26, p_A(t+5)=0.17$
with a continuous 
decline in A vote intentions. The variation of $p_A(t+1)$ as function of $p_A(t)$ is shown in Figure (\ref{p3t})

\begin{figure}[t]
\begin{center}
 \includegraphics[width=.75\textwidth]{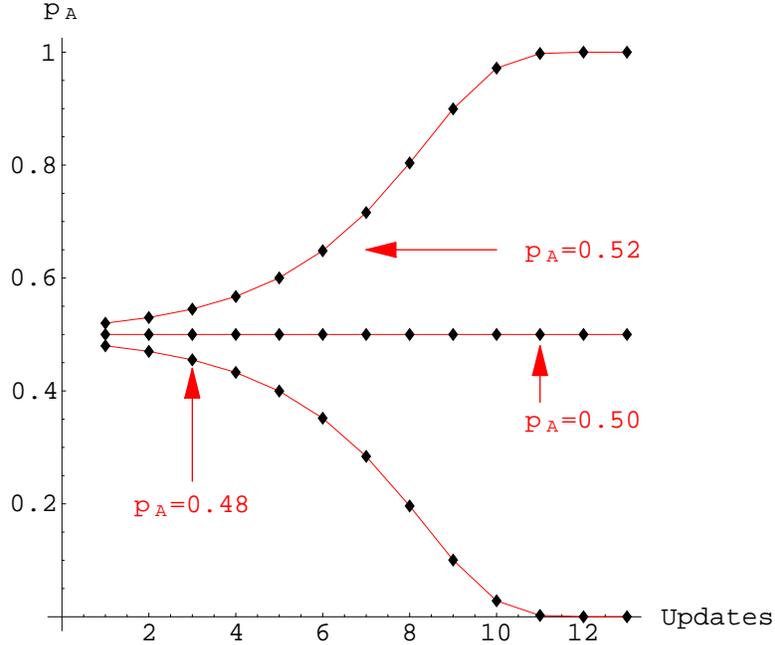}
\caption{
Variation of $p_A(t)$ for groups of size 3 as function of repeated updates with three initial support $p_A(t)=0.48, 0.50, 0.52$. The resulting extremism is democratic since it is along the initial majority.}
\label{p3-updates}
\end{center}
\end{figure}

Adding 3 more cycles would result in zero A vote 
intention with $p_A(t+6)=0.08, p_A(t+7)=0.02$ and $p_A(t+8)=0.00$. Given any initial 
intention vote distribution, the random local 
opinion update leads toward a total polarization of the collective opinion. 
Individual and collective opinions stabilize simultaneously along the same and 
unique vote intention either A or B, depending on the initial support distribution. The evolution of respectively $p_A=0.48, 0.50, 0.52$ as function of repeated updates is exhibited in Figure (\ref{p3-updates})

The update cycle number to reach either one of the two stable attractors can be 
evaluated from Eq. (\ref{p3}). It depends on the distance of the initial densities from 
the unstable point attractor. An approximate analytic formula can derived,

\begin{equation}
n\simeq \frac{1}{\ln[\frac{3}{2}]}\ln[\frac{1}{\mid 2p_A(t)-1\mid}]+1.85,
\label{n}
\end{equation}
where last term is a fitting correction \cite{mino}. The number of 
cycles being an integer, its value is obtained from Eq. (\ref{n}) rounding to 
an integer. It is found to be $n$ always a small number 
as shown in Figure (\ref{n}). Eq. (\ref{n}) yields $8$ at an initial value $p_A(t)=0.45$ 
and $4$ at $p_A(t)=0.30$,
which are the exact values obtained by successive iterations from Eq. (\ref{p3}). 

However, in real life discussions, people do not change their mind at each gathering,
consequently, every one of our formal update cycle takes indeed some time length, which in turn is equivalent to some number of days whose exact evaluation is out the scope of the present work.
 
Therefore, in practical terms the required time to eventually complete the 
polarization process is much larger than the campaign duration, thus preventing 
to actually reach one attractor with a total opinion polarization. From above example at $p_A(t)=0.45$, two cycles yield a result of 
$39\%$ in favor of A and $61\%$ in favor of B. One additional update cycle makes 
$34\%$ in favor of A and $66\%$ in favor of B.

A generalization to any distribution of group sizes was achieved 
yielding a very rich and complex phase diagram \cite{mino}. A unifying frame was found with the property to incorporate all opinion dynamics model \cite{uni}.

\section{The contrarian effect}
 
We can now introduce agents who may become contrarians. A contrarian  is someone who deliberately decides to oppose the prevailing 
choice of others once a local group has reached the consensus driven by the majority rule. The shift is independent of the choice itself \cite{contra}. Contrarian strategy seems to be a growing new trend of modern democracies and was first studied in finance \cite{contra-eco}. 
 
It is worth to stress we are considering the contrarian behavior as a probabilistic attitude. Any agent may adopt it during one update cycle beside its regular local majority rule mechanism. 
Setting contrarian 
choices at a density $a$ with $0\leq a< \frac{1}{2}$, the density of A opinion
given by Eq. (\ref{p3}) 
becomes, 
\begin{equation}
p_A(t+1) =(1-a)[p_A(t)^3+3p_A(t)^2p_B(t)]
+a[p_B(t)^3+3p_B(t)^2p_A(t)] ,
\label{p3a1}
\end{equation}
where first term corresponds to the regular update process and second term to contrarian
contribution from local groups where the local majority was in favor of B. The upper limit constraint on the range of $a$ comes from the definition of a contrarian, which is a behavior opposed to the majority behavior, hence the $a<\frac{1}{2}$ restriction.
A formalization in terms of temperature was performed in \cite{contra-esp}. A Monte Carlo simulation did confirm the main features of the model \cite{contra-simu}

\begin{figure}[t]
\begin{center}
 \includegraphics[width=1.0\textwidth]{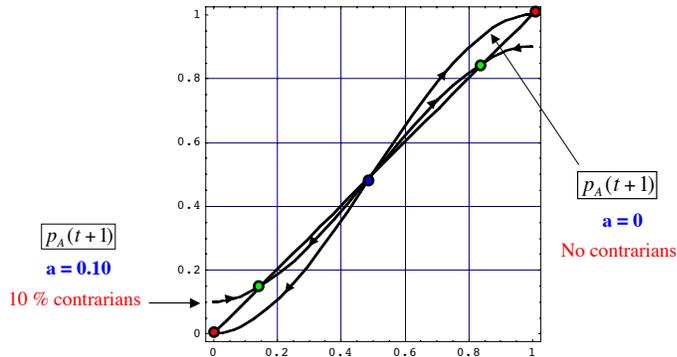}
\caption{Equation (4) with $p_{A}(t+1)$  as function of $p_{A}(t)$ at  a contrarian density of respectively 
$a=0$ and $a=0.10$. 
In the second case the two stable point attractors have moved from total 
polarization towards a coexistence of stable mixed vote intentions with a clear cut 
majority-minority splitting.}
\end{center}
\label{p3a-f}
\end{figure}

Using $p_B=1-p_A$ makes Eq. (\ref{p3a1}) to write
\begin{equation}
p_A(t+1) =  (1-2a)[-2p_A(t)^3+3p_A(t)^2]+a,
\label{p3a2}
\end{equation}
which is shown in Fig. (\ref{p3a-f}) for the case $a=0.10$, i.e., with $10\%$ of contrarian choices as compared to the pure case $a=0$ without contrarians. 

From Eq. (\ref{p3a2}) the effect of low-density contrarian choices is twofold. First both pure stable point attractors from the case $a=0$ are shift toward mixed ones with a
coexistence of vote intentions. As soon as $a\neq 0$ total polarization is averted with for the two symmetric attractors
\begin{equation}
P_{A(B)}=\frac{(1-2a)\pm\sqrt{12a^2-8a+1}}{2(1-2a)},
\label{roots}
\end{equation}
which are defined only in the range  $0\leq a\leq\frac{1}{6}$. It is also worth to notice that   $P_{A}+P_{B}=1$. The index  A(B) means a majority for A(B) with a B(A) minority. For instance a value 
of $a=0.10$ yields $P_{A}=0.85$ and $P_{B}=0.15$. At $P_{A}=0.85$ there  exists 
a stable coexistence of vote intentions at respectively $0.85\%$ in favor of A
with $0.15\%$ for B. The reverse holds at $P_{B}=0.15$. At contrast contrarian 
choices keep unchanged the unstable point separator at $\frac{1}{2}$ whatever is the value of $0\leq a < \frac{1}{6}$.

\begin{figure}[t]
\includegraphics[width=1.0\textwidth]{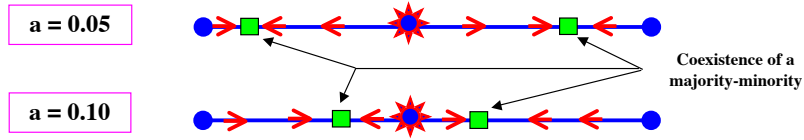}
\caption{
Variation of the flow diagram of opinion dynamics when contrarians are included $(a\neq 0)$ for update groups of size $3$. The cases $a=0.05$ and $a=0.10$ are shown. Arrows show the direction of the flow driven by the public debate for any initial respective supports.}
\label{flow2}
\end{figure}

The second effect from contrarian choices is an increase in the number of 
cycle updates in reaching the stable attractors. For instance starting as above 
at $p_A(t)=0.45$  with $a=0.10$ leads now to the series $p_A(t+1)=0.44, 
p_A(t+2)=0.43, p_A(t+3)=0.42, p_A(t+4)=0.40, p_A(t+5)=0.38$. Additional $12$ 
updates are required to reach the 
stable attractor at $0.15$. All cycles score to $17$ against only $8$ without 
contrarian 
choices. A vote after two update cycles from above same example would give a voting 
result of $43\%$ in favor of A and $57\%$ in favor of B instead of respectively $39\%$ 
and $61\%$ at $a=0$.

As for the non contrarian case an  approximate formula can be derived from Eq. (\ref{p3a2}) to evaluate the update cycle number
required to reach either one of the two stable attractors. It writes, 
\begin{equation}
n\simeq \frac{1}{\ln[\frac{3}{2}(1-2a)]}\ln[\frac{2p_A-1}{\mid 2p_A(t)-1\mid}]
+\frac{1.85}{(1-2a)^{5.2}} ,
\label{na}
\end{equation}
where we have used the property  $P_{A}+P_{B}=1$ and last term is a fitting correction \cite{mino}. At $a=0.10$ we found the exact values of 17 and 9 for respectively $p_A(t)=0.45$ and $p_A(t)=0.30$.

\begin{figure}[t]
\begin{center}
\includegraphics[width=.75\textwidth]{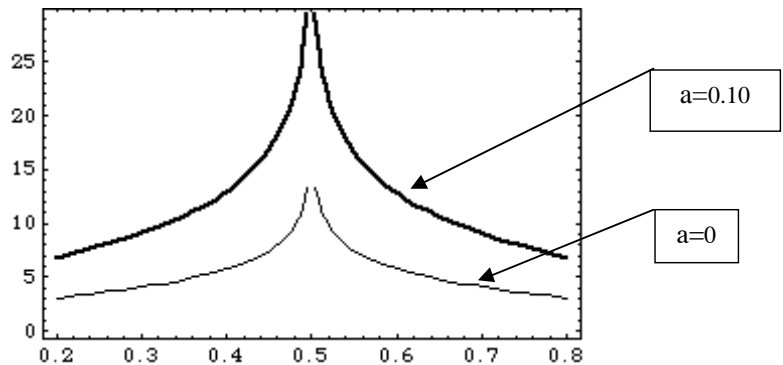}
\caption{Approximate number of cycles of vote intention updates to reach a total 
polarization of opinion as function of an initial support  $p_{A}(t)$.}
\label{n0-na}
\end{center}
\end{figure}

Both Eq. (\ref{na})  and Figure (\ref{n0-na})  show explicitly that the existence of contrarian behavior has a drastic 
effect in increasing the number of required update cycle to reach the stable point 
attractors (which exist only for $a<\frac{1}{6}$). That means much more weeks in a real debate time scale. In practical terms it implies 
a quasi-stable coexistence of both vote intentions not too far from fifty 
percent but yet with a clear-cut majority in one direction, which is determined 
by the initial majority.

\begin{figure}[t]
\includegraphics[width=1.0\textwidth]{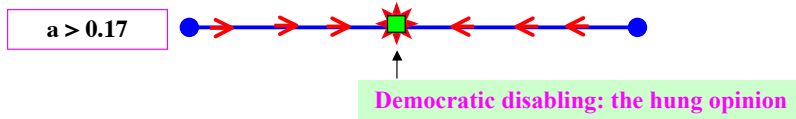}
\caption{
Hung voting fow diagram of opinion dynamics when contrarians are included with $(a>a_c=\frac{1}{6})$ for update groups of size $3$. Arrows show that from any initial supports the public debate drives the opinion distribution toward $\frac{1}{2}$.}
\label{flow3}
\end{figure}

However contrarian choices may lead to an additional radical qualitative change in the 
whole vote intention dynamics. Eq. (\ref{roots}) shows that at a density of 
$a=\frac{1}{6}\simeq 0.17$, contrarian 
choices make both point attractors to merge simultaneously at the 
unstable point separator  $p_c=\frac{1}{2}$ turning it to a stable point attractor as seen in Figure (\ref{flow3})

Consequences on the vote intention dynamics are drastic. The flow 
direction is now reversed making any initial densities of vote intentions to converge via the public debate 
toward a perfect equality between vote intention for A and B. In physical terms, 
contrarians produce a phase transition from an ordered symmetry broken majority-minority phase into a disordered symmetrical fifty 
percent balance phase with no majority-minority splitting. 
In the ordered phase elections always yield a clear-cut majority.

\begin{figure}[t]
\begin{center}
\includegraphics[width=1.0\textwidth]{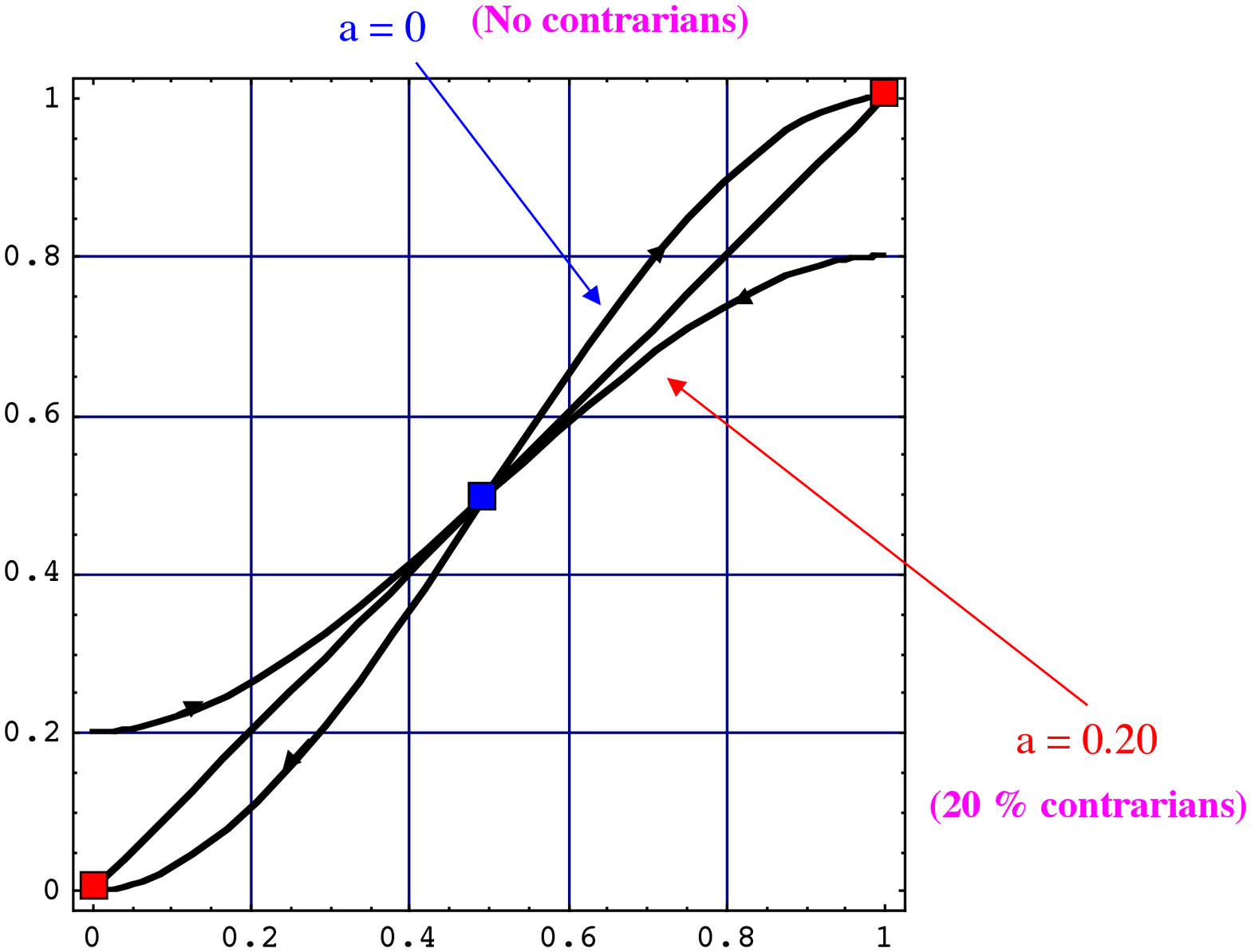}
\caption{$p_{A}(t+1)$ as function of $p_{A}(t)$ at $a=0$ and $a=0.20$. In the first 
case the vote intention 
flows away from the unstable point attractor at $\frac{1}{2}$ toward either one of the stable point attractors at zero or one. In the second case, contrarian choices have 
reversed the flow directions  making any initial densities to flow 
toward $\frac{1}{2}$, now the unique and stable point attractor.}
\end{center}
\label{rev}
\end{figure}    

At contrast in the disordered phase elections lead 
to a random outcome driven by statistical fluctuations with a result very close to fifty-fifty.
An illustration is shown in Figure (8)  for $20\%$ of contrarians.

\section{Group size increase}

In real social life people donÕt meet only by group of 3. However, generalizing 
above approach to larger sizes is straightforward and does not change the 
qualitative feature of the model. Dynamics reversal driven by contrarians 
towards the disorder phase with no majority-minority splitting is preserved. 
The main effect is an 
increase in the value of the contrarian critical density at which the phase 
transition occurs. In the case 
of an odd size k, Eq. (\ref{p3a2}) becomes, 
\begin{equation}
p_A(t+1)=(1-2a)
\sum_{i=\frac{k+1}{2}}^k  C_k^i p_A(t)^i p_B(t)^{(k-i)}+a ,
\end{equation}
where $C_k^i\equiv \frac{k!}{(k-i)!i!}$ . The instrumental parameter in determining 
the flow direction and the
associate phase transition is the eigenvalue at the point attractor $p_c=\frac{1}{2}$. 
It is given by,
\begin{equation}
\lambda=(1-2a)\left[\frac{1}{2}\right]^{k-1}\sum_{i=\frac{k+1}{2}}^k (2i-k)C_k^i .
\label{lambda}
\end{equation}

The range  $\lambda >1$ determines an unstable point attractor with an ordered phase 
characterized by the existence of a majority-minority splitting. 
At contrast, $\lambda <1$ makes the 
point attractor stable. The case  $\lambda =1$ determines the critical value of the 
contrarian choice density $a_c$  at which the phase transition occurs. 
From Eq. (\ref{lambda}), we get,
\begin{equation}
a_c=\frac{1}{2} \left( 1-\left[ (\frac{1}{2})^{k-1}
\sum_{i=\frac{k+1}{2}}^k (2i-k)C_k^i\right]^{-1} \right) .
\end{equation}
In the case $k=3$  we recover the above result $a=\frac{1}{6}\simeq 0.17$. We find $a_c=0.33$ at $k=5$ and $0.30$ at $k=9$ with the limit $a_c\rightarrow \frac{1}{2}, k \rightarrow +\infty$.

\section{Conclusion}
We have presented a simple model to study the effect of contrarian choices 
on opinion forming. At low densities $a$ the opinion dynamics leads to a mixed phase with a 
clear cut majority-minority splitting. However, beyond some critical 
density $a_c$, contrarians make all the attractors to merge at the separator $p_c=\frac{1}{2}$. It 
becomes the unique attractor of the opinion dynamics. 

When $a>a_c$ vote intentions flow 
deterministically with time towards an exact equality between A and B opinions. In this new
disordered stable phase no majority appears. Agents keep shifting opinions 
but no symmetry breaking (i.e., the appearance of a majority) takes place.
There an election would result in effect in a random winner due to statistical fluctuations. 
The value of $a_c$ depends on the size distribution of update groups. 

Accordingly, our results shed a totally new light on recent elections in 
America (2000) and Germany (2002). It suggests those ``hung elections" were not 
chance driven. On the opposite, they are a deterministic outcome 
of contrarians. As a consequence, since contrarian
thinking is becoming a growing trend of modern societies, the subsequent
``hanging chad elections'' syndrome was predicted to become both inevitable 
and of a common occurrence in our earlier work \cite{contra}.

In the mean time our prediction has been fulfilled twice with the 2005 Schr\H{o}der-Merkel German elections and  the 2006 Prodi-Berlusconi Italian elections. Even the 2005 Bush-Kerry American presidential elections, which is at odd with above cases (Bush got a 2.5$\%$ margin with Kerry), is part of the same frame. It indeed demonstrates the effect of applying an external pressure to reinitialize the respective proportions of support for A and B. Up to few days ahead of the vote, the polls were showing a definite move toward a hung vote and it is the Bin Laden videotape, released the Friday preceding the election, which has shifted the election vote. Accordingly to our frame, we claim that would the elections have taken place two weeks latter, the Bush-Kerry race would have yield another hung election as in 2000.

Clearly our model opens new and exciting paths to the modeling of opinion forming. Nevertheless it is also a crude simplification of the complicated reality of social phenomena and we should keep in mind that it is only a model, which does not claim to be an exact description of the very reality. Many new features could be incorporated to enrich it  in social features like recently making the density of contrarian to depend on the total densities of agents supporting each coalition  \cite{contra-pre}.

\end{document}